# Spherical and hyperspherical harmonics representation of van der Waals aggregates

Andrea Lombardi[1,a)], Federico Palazzetti[1], Vincenzo Aquilanti[1], Gaia Grossi[1], Alessandra F. Albernaz[2], Patricia R. P. Barreto[3] and Ana Claudia P. S. Cruz[3]

[1]Dipartimento di Chimica, Biologia e Biotecnologie, Università di Perugia, Via Elce di Sotto 8, 06123, Perugia, Italy.
[2]Instituto de Física, Universidade de Brasília, CP04455, Brasília, DF, CEP 70919-970, Brazil
[3]Instituto Nacional de Pesquisas Espaciais (INPE)/MCT, Laboratòrio Associado de Plasma (LAP), São José dos Campos, SP, CEP 12247-970, CP515, Brazil
a)Corresponding author: andrea.lombardi@unipg.it

**Abstract.** The representation of the potential energy surfaces of atom-molecule or molecular dimers interactions should account faithfully for the symmetry properties of the systems, preserving at the same time a compact analytical form. To this aim, the choice of a proper set of coordinates is a necessary precondition. Here we illustrate a description in terms of hyperspherical coordinates and the expansion of the intermolecular interaction energy in terms of hypersherical harmonics, as a general method for building potential energy surfaces suitable for molecular dynamics simulations of van der Waals aggregates. Examples for the prototypical case diatomic-molecule–diatomic-molecule interactions are shown.

## INTRODUCTION

Spherical and hyperspherical harmonics expansions [1] have been successfully employed to represent potential energy surfaces of a large series of van der Waals aggregates (e. g. floppy molecule-rare-gas-atom [2, 3, 4], $H_2O$–$H_2$ [5], $H_2O$–rare-gas-atom [6]), using information from molecular beam experiments and/or quantum chemical calculations. The general method is based on the expansion of the interaction potential in terms of hyperspherical harmonics and consists in fitting a certain number of points of the potential energy surface (PES), previously determined, selected on the basis of geometrical and physical characteristics of the system. The resulting potential energy function is suitable to serve as a PES for molecular dynamics simulations [7], because of its compact analytical form, allowing for fast calculation of derivatives, and its full account of the symmetries of the system. Here we discuss the spherical and hyperspherical representation of the PESs for the general diatom–diatom case [8], starting from the simple atom-diatom case. We limit our consideration to non-reactive interactions, where the constraint on interatomic bonds and bond angles kept "frozen" at their equilibrium positions is enforced, i. e. the interacting molecules are considered as rigid. Exceptions are those of larger molecules, such as $H_2O_2$ [2], where the "floppy" torsional mode is active. The more complex case of reactions, not explicitly treated here, implies the release of the internal constraints on bonds and angles (see e.g. [9, 10] and a careful choice of the set of angular coordinates among the alternative variants of the hyperspherical ones[11, 12, 13, 14], to get a convenient form of the kinetic energy and of the potential energy surface. The internal dynamics of many-body systems, such as atomic and molecular clusters, is another interesting field of application of this approach, where the adoption of the hyperspherical coordinates (here illustrated in their use for the PESs) permits one to separate the different contributions to the kinetic energy, on the basis of an hyperspherical mode analysis [15, 16, 17, 18, 19, 20, 21]. Variants of the hyperspherical expansion methods have also been applied to the modeling of the intermolecular interactions of molecules relevant to atmospheric chemistry [22, 23, 24, 25, 26, 27].

## THEORETICAL BACKGROUND

The intermolecular interactions involving atoms or molecules can be represented as a sum of an isotropically averaged contribution, as typically measurable from scattering of an atom with a hot (i.e. fast rotating) diatom, and an anisotropic term. In the prototypical case of an atom interacting with a diatomic molecule considered as a rigid rotor, the interaction depends on a distance R, usually the distance between the atom and the center-of-mass of the molecule and an angle θ, defined by the direction of R and the molecular axis. The intermolecular potential V (R, θ), where $0 \leq \theta \leq \pi$ admits an expansion in spherical harmonics, simply the Legendre polynomials $P_\mu(\cos \theta)$ (the orthonormal expansion set with the required completeness properties for convergence for a function defined 0-π range), is expressed as:

$$V(R, \theta) = \sum_\mu V_\mu(R) P_\mu(\cos \theta)$$

where μ = 0, 2, · · · (the odd terms vanish by symmetry). The "minimal" expansion in this case is obtained by truncation, V (R, θ) = $V_0$(R) + $V_2$(R) $P_2$(cos θ), where $P_2$(cos θ) = (3 $\cos^2$(θ) − 1)/2. Identifying $V_\parallel$= V (R, 0) and $V_\perp$ = V(R, π/2), one has the relations

$$V_0(R) = \frac{1}{3}(V_\parallel + 2V_\perp) \; ; \; V_2(R) = \frac{3}{2}(V_\parallel - V_\perp) \qquad (1)$$

or inversely $V_\parallel = V_0 + V_2$, and $V_\perp = V_0 - \frac{1}{2} V_2$.

The above outlined representation can be considered as a spherical-harmonics expansion, such as that of Refs. [1, 6, 5, 2]. This approach to the intermolecular interactions focuses on the isotropic $V_0$ term, the leading one for weakly bonded systems.

## Leading Configurations

To define the leading configurations, let us consider the spherical-harmonics expansion of the intermolecular potential energy surface of two diatomic molecules introduced for the $O_2$–$O_2$ system [28, 29]. For a given configuration of the two molecules, treated as rigid rotors, at a certain value of the distance R of their centers of mass, the intermolecular interaction energy V depends on a set of three angular coordinates, denoted as $\theta_a$, $\theta_b$, ranging from 0 to $\pi$, and $\varphi$, ranging from 0 to $2\pi$, as follows:

$$V(R, \theta_a, \theta_b, \phi) = 4\pi \sum_{L_a, L_b, L} V^{L_a, L_b, L}(R) Y_{L_a, L_b}^{L0}(\theta_a, \theta_b, \phi) \qquad (2)$$

where $L_a, L_b = 0, 1, 2, \cdots$ and $|L_a - L_b| \leq L \leq L_a + L_b$ and the angular functions $Y_{L_a, L_b}^{L0}$ functions are bipolar spherical harmonics:

$$Y_{L_a, L_b}^{L0} = \sum_m (-)^{L_a - L} \begin{pmatrix} L_a & L_b & L \\ m & -m & 0 \end{pmatrix} \times Y_{L_a, m}(\theta_a, \phi_a) Y_{L_b, -m}(\theta_b, \phi_b)$$

where the functions $Y_{L_a, m}$ and $Y_{L_b, -m}$ are ordinary spherical-harmonics, the symbol between large parentheses is a 3−j symbol [30] and it holds the inequality $-\min(L_a, L_b) \leq \min(L_a, L_b)$. The radial coefficients $V^{L_a, L_b, L}(R)$ are the "momenta" of the expansion, representing the radial dependence of the different components of the interaction, i.e. dispersion, induction and electron overlap plus electrostatic contributions. Figure 1 shows the set of leading configurations appropriate for characterizing the interactions of pair of diatomic molecules, with the corresponding values of the above mentioned angular coordinates. At any given configuration, the sperical-harmonics in Equation 2 assume fixed values and the potential energy, depending on R, becomes a linear combination of the corresponding momenta $V_{L_a, L_b, L}(R)$. For a value $R_i$ assigned to the distance R, the interaction energy can be estimated from ab initio calculations, for each of the leading configurations, and a set of values for cuts of the potential energy surface is obtained. The obtained values and the corresponding sets of the $V_{L_a, L_b, L}(R)$ radial momenta, form a system of linear equations, to be solved analitically, expressing the radial momenta as a combination of spherical-harmonics. The isotropic radial term, the $V_{000}(R)$ moment of the expansion, is interpreted in terms of the relative contribution of the size-repulsion, induction and dispersion attraction terms, to the intermolecular interaction, plus possibly additional attractive effects (as those due to charge-transfer components), which may not vanish when averaging over all mutual orientations.

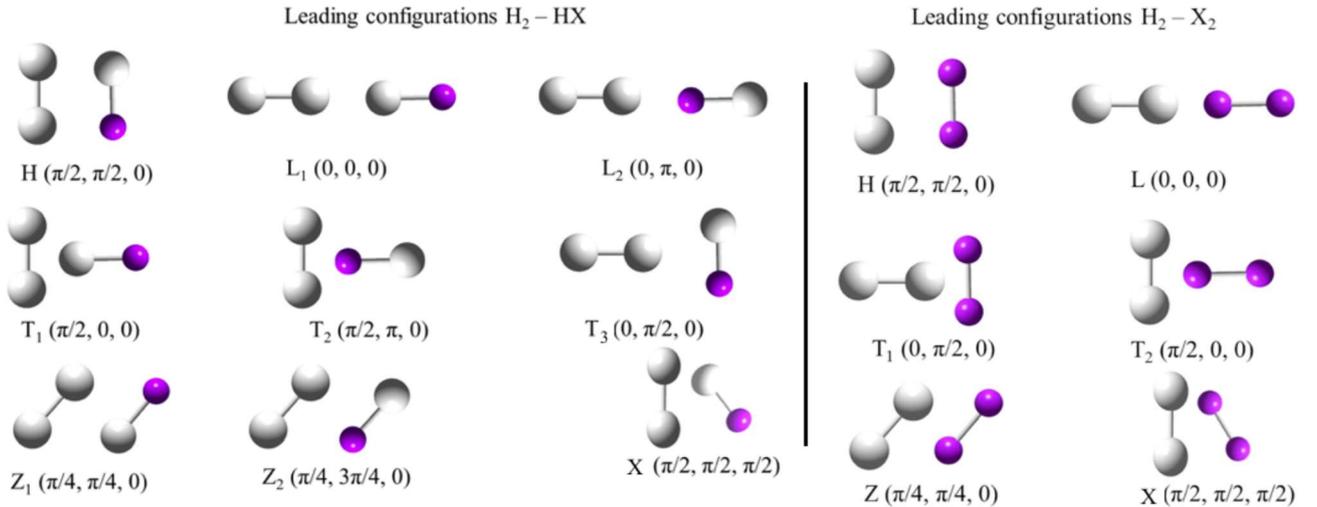

**FIGURE 1**. Sketch of the leading configurations adopted in Ref. [8] to characterize the intermolecular interactions of the $H_2$–$X_2$ and $H_2$–HX systems, X = H, F, Cl. The corresponding values of the $\theta_a$, $\theta_b$ and $\phi$ coordinates are indicated (see Ref. [8]).

# Intermolecular interactions of pairs of diatomic molecules: the H2-X2 case for X = H, F, Cl

The knowledge of the intermolecular interaction between diatomic molecules is of practical and fundamental interest for atmospheric chemistry. In order to provide an example of application of the hyperspherical expansion method to the case of diatomic molecule–diatomic molecule interaction, we report here the isotropic radial term $V_{000}(R)$ of the hyperspherical expansion of the intermolecular potential energy of the $H_2$–$H_2$, $H_2$–$F_2$ and $H_2$–$Cl_2$ pairs of molecules. Figure 2 shows the $V_{000}(R)$ profiles for the three systems.

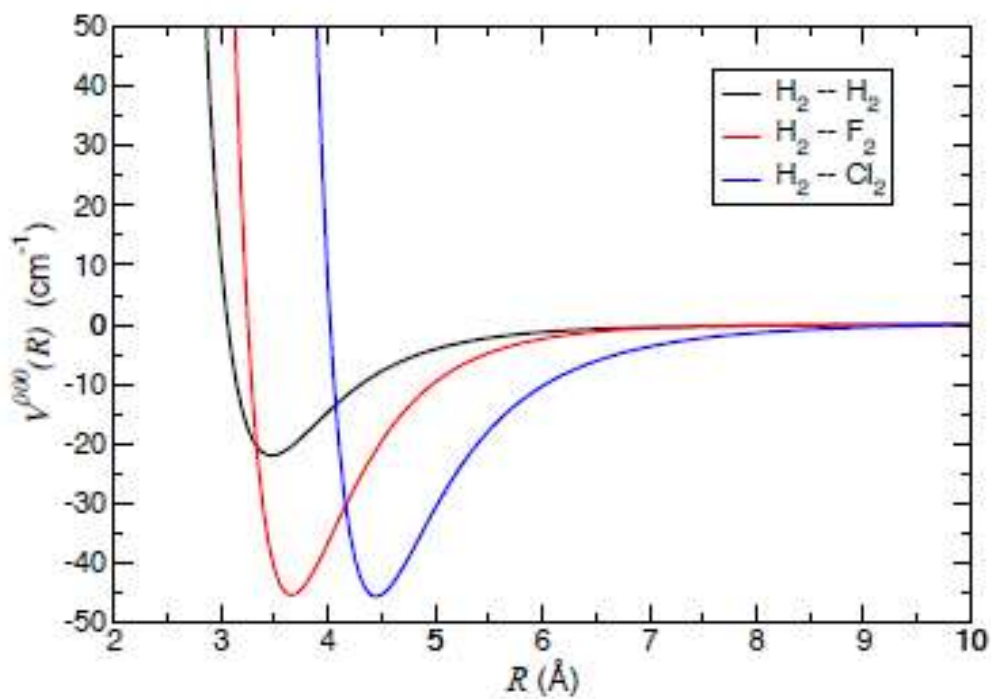

**FIGURE 2.** Isotropic component $V_{000}(R)$ of the intermolecular interactions of $H_2$–$H_2$, $H_2$–$F_2$ and $H_2$–$Cl_2$ systems.

## ACKNOWLEDGMENTS
A.L. acknowledges financial support from the Dipartimento di Chimica, Biologia e Biotecnologie dell'Universit`a di Perugia (FRB,Fondo per la Ricerca di Base), from MIUR PRIN 2010-2011 (contract 2010ERFKXL 002) and from "Fondazione Cassa Risparmio Perugia (Codice Progetto: 2015.0331.021 Ricerca Scientifica e Tecnologica)". F.P., A.L., and V.A. acknowledge the Italian Ministry for Education, University and Research, MIUR for financial support through SIR 2014 Scientific Independence for Young Researchers (RBSI14U3VF) and FIRB 2013 Futuro in Ricerca (RBFR132WSM 003). V.A. thanks CAPES for the appointment as Professor Visitante Especial at Instituto de F´ısica, Universidade Federal de Bahia, Salvador (Brazil).